\documentclass[sigconf,screen,authorversion,nonacm,review=false,timestamp=false]{acmart}

\AtBeginDocument{}

\setcopyright{rightsretained}
\copyrightyear{2024}
\acmYear{2024}
\acmDOI{10.1145/3613904.3642153}
\acmISBN{979-8-4007-0330-0/24/05}

\acmConference[CHI '24]{Proceedings of the CHI Conference on Human Factors in Computing Systems}{May 11--16, 2024}{Honolulu, HI, USA}
\acmBooktitle{Proceedings of the CHI Conference on Human Factors in Computing Systems (CHI '24), May 11--16, 2024, Honolulu, HI, USA}

\usepackage{tabu}
\usepackage{ccicons}
\usepackage{csquotes}
\usepackage{soul}
\usepackage{multirow}

\begin{document}

\title{Capturing Cancer as Music: Cancer Mechanisms Expressed through Musification}

\author{Rostyslav Hnatyshyn}
\affiliation{
\institution{Arizona State University}
\city{Tempe}
\state{Arizona}
\country{USA}}
\email{rhnatysh@asu.edu}

\author{Jiayi Hong}
\affiliation{
\institution{Arizona State University}
\city{Tempe}
\state{Arizona}
\country{USA}}
\email{jhong76@asu.edu}

\author{Ross Maciejewski}
\affiliation{
\institution{Arizona State University}
\city{Tempe}
\state{Arizona}
\country{USA}}
\email{rmacieje@asu.edu}

\author{Christopher Norby}
\affiliation{%
  \institution{Arizona State University}
  \city{Tempe}
  \state{Arizona}
  \country{USA}}
\email{cnorby@asu.edu}

\author{Carlo C. Maley}
\affiliation{%
  \institution{Arizona Cancer Evolution Center, Biodesign Institute, Arizona State University}
  \city{Tempe}
  \state{Arizona}
  \country{USA}}
\email{maley@asu.edu}

\renewcommand{\shortauthors}{Hnatyshyn et al.}
\newcommand{\eg}{e.\,g.}
\newcommand{\ie}{i.\,e.}
\newcommand{\later}[1]{\textcolor{orange!60!black}{#1}}
\newcommand{\jh}[1]{\textcolor{orange}{#1}}
\newcommand{\ross}[1]{\textcolor{blue}{#1}}
\newcommand{\redout}[1]{}
\newcommand{\rev}[1]{#1}

\begin{abstract}
The development of cancer is difficult to express on a simple and intuitive level due to its complexity. Since cancer is so widespread, raising public awareness about its mechanisms can help those affected cope with its realities, as well as inspire others to make lifestyle adjustments and screen for the disease. Unfortunately, studies have shown that cancer literature is too technical for the general public to understand. We found that musification, the process of turning data into music, remains an unexplored avenue for conveying this information. We explore the pedagogical effectiveness of musification through the use of an algorithm that manipulates a piece of music in a manner analogous to the development of cancer. \rev{We conducted two lab studies and found that}\redout{~We conducted a lab study with 13 participants and found that the approach was positively received. The participants found it enhanced their understanding of the mechanisms of cancer when coupled with an article that provides context.}~\rev{our approach is marginally more effective at promoting cancer literacy when accompanied by a text-based article than text-based articles alone.}
\end{abstract}



\maketitle

\section{Introduction}
Cancer is a terribly widespread disease that has affected millions of people. However, certain lifestyle choices have been shown to significantly decrease the risk of developing some types of cancer~\cite{Colditz.2006.EICa}. As a result, increasing public awareness can be highly effective in reducing cancer rates~\cite{Colditz.2006.EICa, Kessler.2017.CCP,Britt.2020.KSE}. Unfortunately, most literature that presents the underlying biological mechanisms of the disease has been found to be difficult for the general public to understand~\cite{Champion.2019.SAM, Chelf.2001.CPE}.

In their work, Champion et. al~\cite{Champion.2019.SAM} suggest an array of alternative methods, including computer-assisted learning as well as multi-media formats. Kemp et al.~\cite{Kemp:2021:HLD} demonstrated that digital health-care approaches increase access to health-care and overall health literacy. Ryhanen et al.~\cite{Ryhanen.2010.EII} found that computer-aided education systems targeting cancer increased cancer literacy rates and encouraged early screening for the disease. Despite \redout{the} evidence that these systems are highly effective in increasing cancer-related literacy and preventative treatment rates, there remains a considerable gap in cancer-focused health technologies~\cite{Prochaska:2017:SMM}; this is due to the complexity of capturing cancer. 

Cancer's complexity arises from its tendency to evolve over time and its unpredictable responses to therapy~\cite{Greaves2012-vo}. Although it is tempting to think of cancer as a unitary thing, any given cancer is made up of billions of cells, carrying millions of mutations, all of which are constantly reproducing, mutating and dying~\cite{Fortunato2017-py}. This dynamic behavior is difficult to capture with static graphics and charts, commonly found throughout cancer literature. We believe that cancer mechanisms may be better illustrated with sound through the process of \textit{Musification}.

Musification, a subset of sonification, is the process of transforming data into \textbf{music}, whereas sonification transforms data into \textbf{sound}~\cite{BonetFilella:2019:DSC}. Sonification has been employed in the design of many diverse applications, from visualizing protein interactions~\cite{Dunn.1999.LMS} to manufacturing process monitoring ~\cite{Hildebrandt.2014.SDS}, as well as countless others~\cite{Sobliye.2017.ULU,E.2006.MNU,Zhou.2004.3SC,Heuten.2006.I3S,Ziemer.2019.PSU}. Sound can be used to describe temporal events and other features of data-sets that are otherwise difficult to visualize~\cite{Sawe.2020.UDS}. This is due to the fact that the ear is better at distinguishing patterns and changes in temporal sequences than the eye~\cite{Guttman.2005.HWE}. These properties make sound an ideal candidate for expressing the complex mechanisms of cancer.

We propose a method to musify cancer by \textit{mutating} a piece of music stochastically, similar to the manner cancer develops in the body. We map \textit{mutations} in cells to operations performed on groups of notes (\ie, measures) within a piece of music. Once the process of mutating the music is complete, a process analogous to \textit{treatment} can be applied. 

Cognizant of the shortcomings of scientific musifications outlined by Williams et al ~\cite{Williams:2016:UCB}, we do not seek to build a totally faithful recreation of cancer, but rather to provide a broad outline that seeks to\redout{~non-expert audiences} \rev{reinforce concepts from an accompanying source}. \redout{Our approach does not provide information but reinforces the existing knowledge of a listener.} 

We evaluated our method's effectiveness in expressing cancer mechanisms through \redout{a study}~\rev{two studies}.~\rev{The results of these studies demonstrate that our approach has a synergistic effect when combined with text-based articles to promote cancer literacy. The resulting piece of music makes the mechanisms of cancer accessible to the general public and provides an additional means for healthcare professionals to explain the disease to patients.} 
\rev{O}ur work contributes the following:
\begin{itemize}
    \item A novel algorithmic approach that simulates the growth of cancer tumors and their treatment through musification;
    \item \redout{A study}~\rev{Two studies} that evaluate\redout{s} the effectiveness of the approach in conveying cancer mechanisms;
    \item A software implementation of the algorithm.
\end{itemize}  

\section{Related Work}
In this section, we explore approaches to cancer education as well as ways musification and sonification have been used in educational and practical contexts.

\subsection{Digital Approaches to Cancer Education}
Most digital approaches to cancer education focus on interactive audio-visual displays to convey information about cancer~\cite{Ryhanen.2010.EII}. Acuna et al.~\cite{Acuna.2020.HDV} surveyed the literature for cancer education videos and found that they were effective in increasing understanding and awareness of cancer mechanisms, as well as increasing screening rates. Jibaja et al.~\cite{Jibaja.2000.TIS} staged an interactive soap opera to better serve Spanish speaking Hispanic women. They found participants gained a stronger understanding of cancer and were more likely to screen for the disease. Ozanne et al.~\cite{Ozanne.2007.PTC} built a complex informational system to aid patients in decision making about their disease with similar positive results. Schonborn et al.~\cite{Schonborn.2016.NEI} developed a visual analytics system that explained the complicated inter-molecular dynamics at play between cancer cells and methods of nano-treatment. However, this system focuses solely on the nano-scale details of cancer, and not on the effects of cancer and its treatment on the body as a whole. 

\rev{A large body of work focuses on the development of interactive video games for promoting cancer literacy among patients~\cite{DeLaHeraConde-pumpido.2018.PRD}. A notable example is Re-mission~\cite{Beale.2007.ICK}, an action game developed for young cancer patients. A longitudinal study demonstrated that patients had improved levels of cancer literacy after interacting with the game for a few hours, leading to improved patient outcomes and quality of life. Gerling et. al~\cite{Gerling.2011.DEC} designed an educational game called Cytarius and ran studies demonstrating that it promoted cancer literacy and elicited positive reactions from patients. These works indicate that innovative interactive methods that promote cancer literacy are highly effective and are associated with improved patient outcomes. We position our work within this area as a yet unexplored alternative to video game technologies. While these technologies are highly effective, they can be inaccessible to those with visual and motor impairments; moreover, the side-effects of cancer treatment can make it difficult for patients to focus on highly stimulating activities such as playing games~\cite{Altun.2018.MCS}.} 

\subsection{Musification}
Musification goes beyond sonification by mapping data to musical structures through the inclusion of tonality and rhythm~\cite{coop2016sonification}. Godbout et al.~\cite{Godbout:2018:EM} built an emotional musification system using the reactions of players playing a music-based game. As players interact with the game, the music changes based on the emotional state of the player which is measured with heart-rate and skin conductance monitors. Once they gathered enough data, they stopped measuring these physical characteristics and instead matched the state of the game world to a player's emotional state. Mendoza et al.~\cite{mendoza2022musification} built a system that would musify accelerometry data from a data-set. The resulting piece of music would be a record of a subject's physical activities over the course of a day. Ultimately, these records were shown to participants in a study to raise awareness about the importance of physical activity. In the study, participants correctly identified the subject with the most physical activity due to the structure of the piece they generated. Mainka et. al~\cite{Mainka.2021.PMS} developed a musification that would provide feedback on the movement of a participant's arms to improve their gait, specifically designed for people suffering with Parkinson's disease. Mcgee et al.~\cite{mcgee2016musification} musified seismic data, generating pieces of music corresponding to the movement of earthquakes.

Unfortunately, most musifications do not sound pleasing due to the nature of the data they present. Williams et al.~\cite{Williams:2016:UCB} argue that there is an inherent trade-off between the accuracy of a musification and its aesthetics. This is due to the inherent chaos present in all data-sets as well as our own cultural biases of what music ought to be. This idea is supported by Coop's framework for sonification~\cite{coop2016sonification}; they consider the musicality of a signal to be a function of its organizational structure that emerges from the stochasticity of a process. In contrast, our approach presents the inverse of this idea - we start from music, and apply the process we are interested in modelling. 

\subsection{Education with Sonification}
Sonification has successfully been applied in a number of educational contexts. Sanchez et al.~\cite{Sanchez.2009.SLB} built a game-like system for teaching visually-impaired students scientific concepts through audio. Similarly, Tomlinson et al.~\cite{Tomlinson:2020:ADI} taught students about Ohm's law and resistance in wires through a system using visualizations coupled with sonification. xSonifiy~\cite{Candey.2006.XS} targeted visually-impaired astrophysicists and students by delivering sonifications of space data and facilitating various interactions with graphs through sound. Adams et al.~\cite{Adams.2022.SSA} developed a library to simplify the sonification of algorithms and included several examples of sonified sorting algorithms. While these approaches focus on visually-impaired audiences, many studies have found that even non-visually impaired students can benefit from sonification.
For instance, Zanella et al.~\cite{Zanella.2022.SSD} explored applying sonification to astronomical data. They argue that sound is ideal for expressing multi-dimensional data due to the various parameters that can be exploited, i.e., pitch, volume, tempo, timbre, etc. Scaletti et al.~\cite{Scaletti.2022.SLM} taught undergraduate students the phenomena of protein folding using a sound-enhanced animation. They found that sonification strongly complemented their visualization, and make it possible to express concepts that were difficult to visualize; in this case, they sonified changes in the conformations of the protein molecules.  Matinfar et al.~\cite{Matinfar.2023.SRA} conducted a study where a surgical assistance system was replaced with an aural alternative. The study found that surgeons were just as accurate with the sonification system and did not need to re-direct their focus from their surgical tools to the visual display. Newbold et al.~\cite{Newbold:2016:MIS} built a sonification system for facilitating recovery progress in patients with chronic pain. Their system would use sound to inform patients on safe boundaries for stretching, using the concept of musical stability within chord progressions to indicate when they approached a point which would cause pain. Musical stability refers to the feeling that a song has concluded; signaled by a cadence, a series of chord changes that ``resolve'' a piece of music. Generally, sonification can strongly enhance existing visualizations and systems, even for those without visual impairments.

\section{Background}
\begin{figure}
    \centering
    \includegraphics[width=\columnwidth]{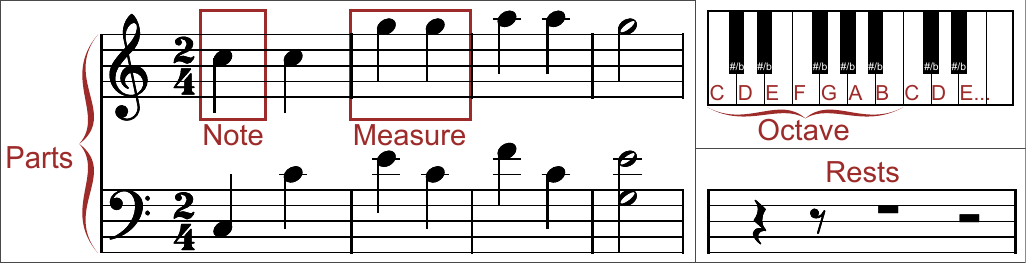}
    \caption{A brief illustration of musical terms necessary to understand our approach. Instruments within a piece are assigned \emph{parts}, which are split into groups of \emph{notes} called \emph{measures}. A \emph{note} informs the musician which sound should be made on their instrument, while a rest denotes silence. At the top right, a piano keyboard is shown to illustrate the concept of \emph{tones} and \emph{octaves}. Each key plays a tone that resides within an octave -- there are twelve tones within an octave, and they repeat indefinitely.}
    \label{fig:terms}
\end{figure}
In this section, we explain various musical concepts that inform our approach, as well as basic information on biology and the mechanisms of cancer. 

A piece of music is typically split up into one or more \emph{parts}, where each part is played by a different instrument. 
It should be noted that there are instruments that can play more than one part at a time, such as the piano. 
Each part is split up into \emph{measures}, which contain a number of \emph{notes} arranged in the order they are played in (Figure~\ref{fig:terms}).
Notes have a \emph{pitch} and a \emph{duration}. 
The pitch of a note is used to determine what frequency should be played on the instrument; in standard western music, the frequency spectrum is divided into \emph{octaves}, with each octave containing 12 unique \emph{tones}. Silence (the absence of a note being played) is denoted by rests, which only have a duration.
The duration of each note or rest is counted in divisions of \emph{beats}, where a \emph{whole} note gets 4 beats, a \emph{half} note is played for 2 beats, a \emph{quarter} note is played for 1 beat, and so on. 
For a complete treatment of music theory, we refer readers to ~\cite{clendinning:2016:musician}.

\emph{Cells} are the basic building blocks of every living organism. Healthy human cells contain copies of DNA - a molecule that contains genetic information for the development and functioning of an organism. Healthy cells grow by dividing and copying their DNA over to their daughter cells which occurs indefinitely in the \emph{cell cycle}. Cancer distorts the DNA within a cell, which causes it to behave incorrectly. When cancer develops in a cell, the cell stops doing the work of its organ and instead starts dividing out of control. The accumulation of abnormal cells in tissue eventually forms a tumor which can then spread to other organs within the body~\cite{WIC.2007.WCN}.

\section{Capturing Cancer with Music}
We collaborated with cancer experts to develop a metaphor for expressing cancer mechanisms through music. \rev{Our principal expert has over 18 years of experience working in computational biology. We also collaborated with a number of doctors from local hospitals during the initial design of our approach.} We met regularly to provide progress updates, solicit feedback and refine the design iteratively.

\rev{Our design was inspired by the inherent parallels between sheet music and DNA. Both DNA and sheet music are simply sequences of instructions -- sheet music provides instructions on how to perform a piece, while DNA provides instructions on how cells function. From a very simplified point of view, cancer is simply a disease that changes the DNA of cells within the body. Therefore, manipulating music to represent cancer is a natural analogy.}

\subsection{The Analogy: Representing Cancer in Music}
A multipart piece of music is like a multi-cellular organism. In this analogy, each part is like the different cell types in a body. So perhaps the violins are the skin cells, the trumpets are the lung cells, the oboes are the liver cells, and so on, with the different timbres distinguishing the different cell types. 

 We represent the start of cancer in a musical piece by randomly selecting one of the parts, starting at a random point near the beginning of the piece. The following $n$ measures of that part become a cancer theme, or \emph{leitmotif}, with $n$ being an adjustable parameter. Instead of that part progressing past those $n$ measures, it starts to repeat that leitmotif, over and over again (Figure~\ref{fig:action}). This represents the cell no longer doing the work of the organ and \emph{dividing out of control}. Each repeat of the cancer leitmotif is like the cancer cell going through the \emph{cell cycle} and reproducing. Each time the cancer theme progresses through its measures, it spawns a copy of itself. Because this represents a \emph{cell dividing}, producing two daughter cells, both the original cancer part and the new cancer part may mutate when they reproduce.

\begin{figure}
    \centering
    \includegraphics[width=\columnwidth]{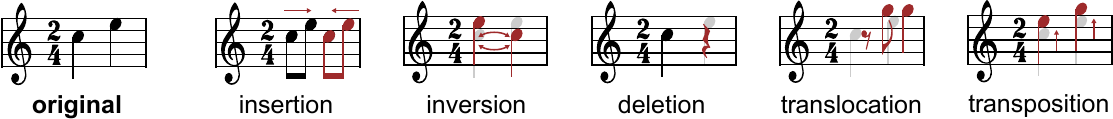}
    \caption{Examples of possible mutations within a piece. Going from left to right, insertion takes a sub-sequence of notes and repeats it (note that the C and E become CECE). Inversion reverses the order of the notes. Deletion replaces a note with a rest. Translocation replaces the measure with a random measure from the rest of the piece. Transposition modifies the pitch of the notes in the measure by a random amount.}
    \label{fig:mutations}
\end{figure}

The DNA in cancer cells \emph{mutates} upon division~\cite{Greaves2012-vo}. We represent these mutations by changing the notes in the musical measures of the cancer leitmotif (see Figure~\ref{fig:mutations}):
\begin{itemize}
    \item An \emph{insertion} in which a short sequence of DNA is repeated, is represented by subdividing a randomly selected series of notes and repeating them. For instance, a set of notes ABC would have their durations divided in half and repeated to form ABCABC. The note durations need to be modified in order to ensure that all measures are of uniform length.
    \item An \emph{inversion} in which a sequence of DNA in the cell is reversed, is represented by randomly selecting a sequence of notes in the cancer leitmotif and reversing the order of those notes.
    \item A \emph{deletion} in which a short sequence of DNA is removed is represented by changing a randomly selected series of notes in the leitmotif into rests.
    \item A \emph{translocation} occurs when DNA breaks and then is repaired by fusing it to another bit of DNA in the cell. In this way, the start of one gene can be fused to the end of another gene, generating an entirely new gene. We represent this by randomly choosing a measure of the cancer leitmotif and replacing it with another measure selected randomly from elsewhere in the piece.
    \item A \emph{\redout{point mutation}~\rev{transposition}, a.k.a, single nucleotide variation}, in which a single letter (base) in the DNA is changed to a different letter, is represented as a change in pitch of a single randomly selected note in the leitmotif. We randomly select a new pitch with a uniform probability among 12 half-steps above and below the current note.
\end{itemize}

We represent cancer therapy at a particular point in time in the piece by silencing the cancer leitmotifs, i.e., replacing all of its notes with rests. However, therapy often fails because cancer cells can mutate in ways that make them resistant to treatment~\cite{Ramos2015-tp}. To represent this, we randomly select a percentage of mutant parts that survive the therapy. As it continues to repeat its leitmotif and reproduce, the tumor regrows.

\definecolor{red}{HTML}{bc0000}
\definecolor{green}{HTML}{4fb286}
\definecolor{blue}{HTML}{00a5cf}
\definecolor{purple}{HTML}{9b5de5}

\subsection{Implementation}
\begin{figure}
    \centering
    \includegraphics[width=\columnwidth]{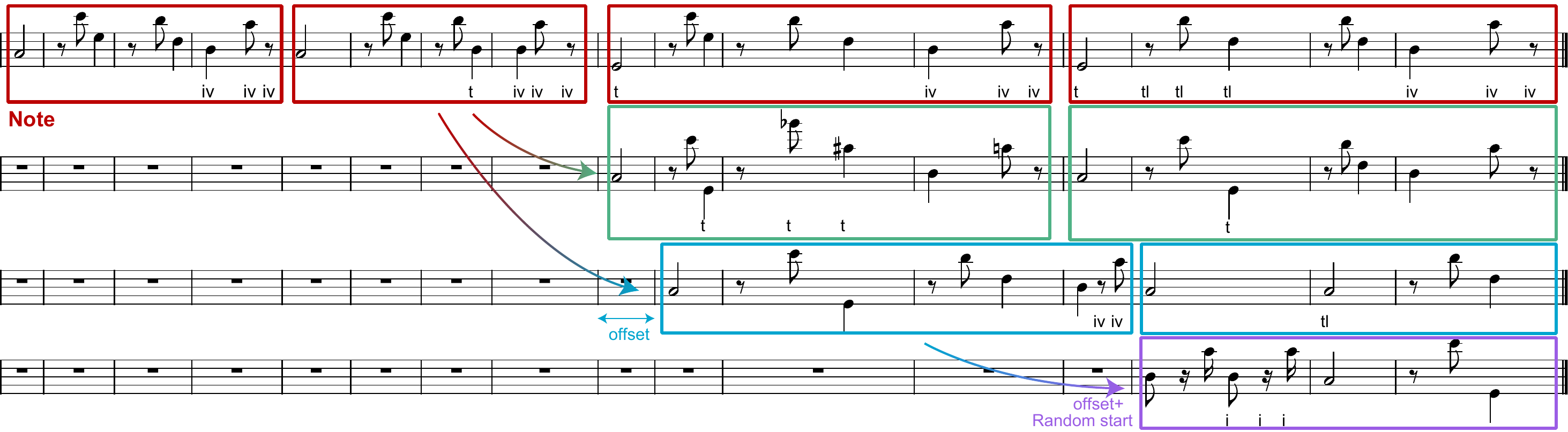}
    \caption{A sample mutation in action. The boxes denote the cancer theme that is mutating. The top-most \textcolor{red}{red} theme was replicated twice, creating two more mutant parts, marked in \textcolor{green}{green} and \textcolor{blue}{blue}. Then, one of the \textcolor{blue}{blue} child mutated again, forming the \textcolor{purple}{purple} part, for three in total. Some of the mutant parts are randomly offset in order to create more variation in the piece, as otherwise the parts would all overlap. }
    \label{fig:action}
\end{figure}
We implemented the algorithm as a Python program that can be run either as an executable, or as a \emph{FastAPI}\footnote{\href{https://fastapi.tiangolo.com}{fastapi.tiangolo.com}} server with a \emph{Svelte}\footnote{\href{https://svelte.dev}{svelte.dev}} front-end. The software takes takes any \emph{MusicXML}\footnote{\href{https://www.musicxml.com}{musicxml.com}} file as input and produces a ``mutant" version of the file as output which allows listeners to mutate any song they wish, provided it is in the MusicXML format. Processing the \emph{MusicXML} file is done with the \emph{music21}\footnote{\href{https://web.mit.edu/music21/}{web.mit.edu/music21}} library. There is also an option to enable treatment through the user-interface or through command-line arguments. Table~\ref{tab:parameters} provides a comprehensive list of options that can be used to modify the output of the algorithm.

\begin{table}[]
\begin{tabular*}{\columnwidth}{p{0.25\columnwidth} | p{0.75\columnwidth}}
\toprule
\textbf{Parameter}                & \textbf{Description} \\\midrule
\emph{Insertion}      & Probability that an insertion will occur.                                      \\
\emph{Deletion}       & Probability that a deletion will occur.                                        \\
\emph{Inversion}      & Probability that an inversion will occur.                                          \\
\emph{Translocation}  & Probability that a translocation will occur.                                      \\
\emph{Transposition}  & Probability that a transposition will occur.                                     \\
\emph{\# of cancer parts} & How many offspring a mutant part can produce.                                                       \\
\emph{Cancer start}               & Percentage relative to the length of the piece.                       \\
\emph{Cancer length}    & Length in measures of the cancer leitmotif.                                                              \\
\emph{Reproduction}   & Probability that a mutant part will reproduce.                                       \\
\emph{Treatment}         & Boolean if treatment should be applied.                                                       \\
\emph{Survival Rate}       & Probability of a mutant to resist treatment.   \\
\emph{Therapy start}              & Percentage relative to the length of the piece.  \\\bottomrule                        
\end{tabular*}
\caption{Parameters available for adjusting the output of the musification algorithm.}
\label{tab:parameters}
\end{table}

\section{Evaluation}
\label{sec:evaluation}
We \rev{initially} evaluated our approach based on three questions:
\begin{itemize}
    \item \textbf{Q1}: Does our approach convey the mechanisms of cancer effectively? 
    \item \textbf{Q2}: \rev{Is a musical background necessary to benefit from our approach?}
    \item \textbf{Q3}: How do listeners react to our approach?
\end{itemize}
 \textbf{Q1} evaluates the effectiveness of our approach; i.e., do listeners gain a \textbf{better understanding} of cancer by listening to cancer-ridden versions of songs they know? \redout{We also ask participants if they believe the music \textbf{reinforces} their understanding of cancer mechanisms.} \textbf{Q2} evaluates whether or not our approach is suitable for audiences without a musical background. \textbf{Q3} evaluates the impact our approach had, as well as its perceived usefulness. \redout{To answer \textbf{Q3}, we conducted a post-study interview and interpreted the responses to each question.}

\subsection{Study Design}
\rev{We conducted two studies to address these questions. The initial study was designed to answer \text{Q1}, \textbf{Q2} and \textbf{Q3} qualitatively. The second study built upon the results and insights of the initial study. The second study aimed to definitively answer \textbf{Q1} and \textbf{Q2} by quantitatively assessing the difference in cancer literacy levels before and after being exposed to our approach. We also included short answer questions in the second study to provide more data for \textbf{Q3}. Both studies were approved by our institution's ethical review board.}

\subsubsection{Initial study}
\rev{Before running the initial study, we realized that context is necessary to understand our approach, as the music does not stand well on its own without some description. Therefore, both studies were designed so that participants would read an article about cancer before being exposed to our approach. The content of the article was derived from the information available on the website of a national cancer-fighting organization (\href{https://www.cancer.org/}{American Cancer Society}). }

\rev{Since our approach \textit{mutates} a piece of music to the point it becomes unrecognizable, we needed to pick a song that would be familiar to most participants. As a result, we chose ``Twinkle Twinkle Little Star'' because it is a well-known lullaby. We generated two 30-second ``mutated'' versions because they showcased each type of mutation without the mutations spiraling out of control. Once the pieces were generated, we put together a video that first played the original version of the song, followed by the two mutant versions.}

\rev{To answer \textbf{Q1} and \textbf{Q2}, we asked participants to rate their understanding of cancer before and after being exposed to our approach. To do so, we developed two 5-point Likert scales, one that assessed a participant's understanding of the mechanisms of cancer (1 meaning the participant had no knowledge and 5 meaning the participant was an expert) and another that assessed the perceived level of knowledge reinforcement our approach provides (1 providing no benefit and 5 providing the most benefit). We also included a Likert scale that assessed a participant's perceived knowledge of sheet music (1 meaning the participant had never seen sheet music before and 5 meaning the participant was comfortable with reading sheet music) (see Appendix~\ref{appendix:a}). Participants were considered to lack a background in music if they reported their musical knowledge as 2 or below. In addition to these scales, we developed a set of questions to assess a participant's attitude towards our approach to answer \textbf{Q3} (see ~\autoref{appendix:b}).}

\rev{\emph{Participants.} We recruited 13 participants (5 female and 8 male) with ages ranging from 20 to 67 (mean=29.69, SD=13.40), denoted through the rest of this work as P1-P13. Participants were solicited from our institution's general engineering Slack channel. One participant had a high-school diploma, seven had Bachelor's degrees and five had Master's degrees. Ten of our participants are university students, with six majoring in computer science, one in robotics, one in biomedical science, and the other two in mechanical engineering. The other participants were a part of the labor force: one works as a user experience designer, one is a semi-retired professor who specializes in cancer research, and one worked as a social worker.}

\rev{\emph{Procedure.} The initial study was performed remotely via Zoom, a conferencing tool, in conjunction with Qualtrics, a survey tool, both of which are hosted by our institution. During the survey, participants were first introduced to the basic mechanisms of cancer through articles based on the information provided in \href{https://www.cancer.org/}{American Cancer Society}. We asked participants to self-report their understanding of cancer. Then, we introduced our approach with a short description that included figures and sample audio demonstrating the various types of mutations. Once the description was read, participants then watched a pre-recorded performance of ``Twinkle Twinkle Little Star'' as well as two mutated versions. Afterward, participants were required to rate their understanding of cancer as well as the perceived utility of the approach. Then, we performed a semi-structured interview. Each interview, on average, took about 30 minutes. Participants were asked to share their screens when they were answering the survey, and we recorded their activities during the survey and their feedback with their permission. At the end of the interview, participants were compensated with a \$15 Amazon gift card.}


\subsubsection{Second study}
\rev{To thoroughly compare our approach to traditional text-based articles, we conducted a follow-up study with objective analysis methods to answer \textbf{Q1} and \textbf{Q2}. We split participants into two groups: Group A which only read the article from the first study as a control group, and Group B which read the article and were exposed to our approach. We asked participants typical demographic questions, as well as a question that asked them to briefly describe their personal experiences with cancer. Both groups were asked a set of cancer literacy questions (\autoref{appendix:c}) prior to reading the article and after they were finished with their tasks. Participants were scored based on the number of correct responses they provided and the two test results were compared to yield their change in cancer literacy levels; scores could range from 0 to 8. To quantify emotional responses, both groups filled out the Self-Assessment Manikin (SAM)~\cite{Bradley.1994.MES} to rate their experience; the SAM is an experimentally verified psychological inventory that measures the emotional impact of a stimulus. Group B also answered a set of questions that were designed to test their knowledge of music theory (\autoref{appendix:d}); Group A did not need to answer these questions because they were not exposed to any music during the study. Participants were scored based on the correct responses they provided to these questions to determine their level of musical knowledge; scores could range from 0 to 5. While listening to the music, Group B participants were asked to mark up a waveform of the audio where they believed the cancer in the song started and when the therapy started working. This question was intended to determine the clarity of our approach. These questions allowed us to measure the emotional and pedagogical impact our approach has compared to a text-based article, as well as determine whether a participant's background could affect their experience.} 

\rev{\emph{Participants.} We recruited 25 participants for both groups, for a total of 50. Participants were recruited entirely from Prolific. Their ages ranged from 20 to 64, with a mean age of 26.6 and a standard deviation of 8.16. 18 participants had earned high school diplomas, 28 participants had earned Bachelor's degrees, 3 participants had earned Master's degrees, and 1 participant had no degrees. 23 participants were university students pursuing degrees in fields ranging from Finance to Biomedical engineering. The other 27 professional participants also had careers in diverse fields, from photography to dentistry.}

\rev{\emph{Procedure.} The study was conducted via Qualtrics. Group A answered a set of demographic questions, then completed the pre-survey cancer literacy assessment, read an article about cancer, filled out the SAM, and then completed the post-survey cancer literacy assessment. Group B additionally answered the musical literacy questions prior to reading the article but otherwise followed the same procedure. After reading the article about cancer, they read a short article that described our approach, and listened to a mutated version of ``Twinkle Twinkle Little Star''. We used the same version of ``Twinkle Twinkle Little Star'' as the initial study, but we repeated the entire piece four times to give participants enough time to fully experience the extent of the mutations. Group B participants then marked up the audio, filled out the SAM, completed the post-survey cancer knowledge assessment and answered a set of short-answer questions about their experience (\autoref{appendix:e}). Participants in group A took an average of 11 minutes to complete the study, while participants in group B took an average of 35.5 minutes. Both groups were compensated at a rate of $\$12$ an hour, with group A earning $\$3$ for 15 minutes of work, and group B earning $\$6$ for 30 minutes of work.}

\section{Results}
\label{sec:results}
In this section, we report our results based on the responses from our participants. \rev{We first interpret the feedback of participants from both studies in a qualitative manner, then analyze the quantitative results of the second study.} \redout{We analyze the results of our Likert scales quantitatively and interpret the feedback of participants in a qualitative manner.}

\subsection{Initial Interview Results}

\textbf{General impressions.} Every single participant described the approach using the words ``interesting'' and ``unique''. Out of \rev{our initial} thirteen participants, twelve had a positive impression, except P2. All of the participants who had a positive impression said they would recommend our approach as an alternative method to teach others about cancer.

However, six participants found the demonstration to be too short; two participants watched the video several times until they felt comfortable. We \rev{initially} chose a short piece because we felt that having a longer piece would cause the music to become overwhelming and difficult to follow due to a large amount of mutations occurring. \rev{This led us to using a longer version in the follow-up study.}~\redout{A longer demonstration would likely have helped participants to clearly understand the approach; despite this, listeners felt that the music reinforced the concepts they learned in the article.}

\textbf{Prerequisites to understanding.} Eight participants said that the article provided the context necessary to understand our approach; without it, the music would be meaningless. On the other hand, three participants said that the context was unnecessary and that a short description would be enough to give future listeners the main idea. Out of these three participants, two watched the video twice to better understand the approach.\redout{, suggesting that a longer video may have been able to stand on its own.}

Ten participants without a background in music said that the cancerous parts and the subsequent therapy were obvious. This is in direct contrast to what the three participants with musical backgrounds believed; participants in this group felt that the music would not help those without experience.

\textbf{Perceived usefulness.} Five participants felt that our approach would be most successful in educational contexts, as the music would appeal to children. These participants felt that the articles we provided in our study were difficult for audiences without an advanced education to comprehend, echoing the results of Champion et al.~\cite{Champion.2019.SAM}. Out of these participants, four directly mentioned that this approach could be useful to those who learn in alternative ways. \textit{P8} said 
\begin{displayquote}
 ``I feel like this approach would be best for those without strong reading abilities, like people who learned English as a second language or children. Music is a universal language.''
\end{displayquote}

\textit{P2}, the cancer expert, did not believe that our approach was useful. They have twenty-five years of experience in the field, and they primarily work with connecting researchers with patients.
\begin{displayquote}
    ``This approach is original and interesting, but I don't think anything about is particularly useful to anyone. It's not realistic enough to be useful for researchers, and I don't think patients care about mutations in their genes. ''
\end{displayquote}
They felt that patients do not care about the particular details of their disease. Furthermore, they felt that our approach was far too unrealistic to be of any educational value. They suggested we should refine our approach and target cancer experts instead by musifying real-world cancer data-sets. However, these impressions were directly contradicted by a participant who had directly experienced cancer. 


\textit{P11}, who is currently living with cancer, had a strong emotional reaction to our approach - they felt we very accurately depicted the feeling of having cancer, but they did acknowledge that the way we implemented therapy was unrealistic. Overall, they felt that our approach was a much better alternative to the print media they had to read, as cancer treatment often has negative cognitive effects on patients.
\begin{displayquote}
``I have chemo-brain\footnote{Chemo-therapy's negative effects on cognition~\cite{Staat.2005.PCB}}. It's nice to just turn my brain off and understand cancer in a different way. I was getting lost in the words while reading the article, but the music spoke to me in a way words can't.''
\end{displayquote} 
Moreover, reading about cancer can be very triggering for patients, making our approach a suitable alternative for exploring the subject of cancer in a non-threatening manner. This sentiment was echoed by \textit{P10} - they found that our approach was unoffensive and ``not intimidating'', unlike the swathes of cancer literature available. P11's strong emotions suggest that our approach has hidden therapeutic potential we did not originally consider in our design.

\textbf{Lessons learned.} Eleven participants felt like they gained a greater understanding of cancer by listening to the music. Seven participants felt that the music provided a general understanding of the mechanisms of cancer and appealed to them on an intuitive and emotional level rather than an intellectual one. This is because the intimate details of each mutation were difficult to follow, as they occur rapidly in great numbers. Three participants explicitly stated the cancer in the piece was obvious to them because the music sounded ``wrong'' or ``weird''. The therapy was equally clear, as the music stopped sounding ``wrong". \textit{P3} and \textit{P4} felt inspired to ``take care of themselves''. \textit{P3} said 
\begin{displayquote}
    ``This reminds me that cancer is very scary. It's important to eat well and live healthily to prevent it. I want to remind my parents to screen for this disease before it's too late.''
\end{displayquote}

\textit{P11} did not feel that they learned much about cancer, but they said this was influenced by their experiences. 
\begin{displayquote}
``I don't feel like this taught me much about cancer. It's not to say it wouldn't; it's just that I've been living with this disease for nearly a decade. I think this would be best for people who are just learning about their diagnosis.''
\end{displayquote} 
This indicates that our tool may best serve newly diagnosed patients, as patients who have been living with cancer for many years will likely have learned what they wanted to know about their disease.

\textbf{Audio-visual comparison.} Eight participants said that the audio was the most helpful element in our approach; the sheet music was inaccessible to them because they could not read it. Surprisingly, two participants without a formal background in music felt that the sheet music made the cancer parts and therapy obvious, even though they could not directly interpret what the notes were. P10 was able to form their own interpretation of the sheet music --
\begin{displayquote}
    ``I don't know what the notes mean, but the structure of each part makes it obvious where the cancer is stronger, and the notes being absent from a part means the cancer cells died. The annotations really help make the mutations clear.''
\end{displayquote}

\subsection{Second Study Results}
\begin{figure}
    \centering
    \includegraphics[width=\columnwidth]{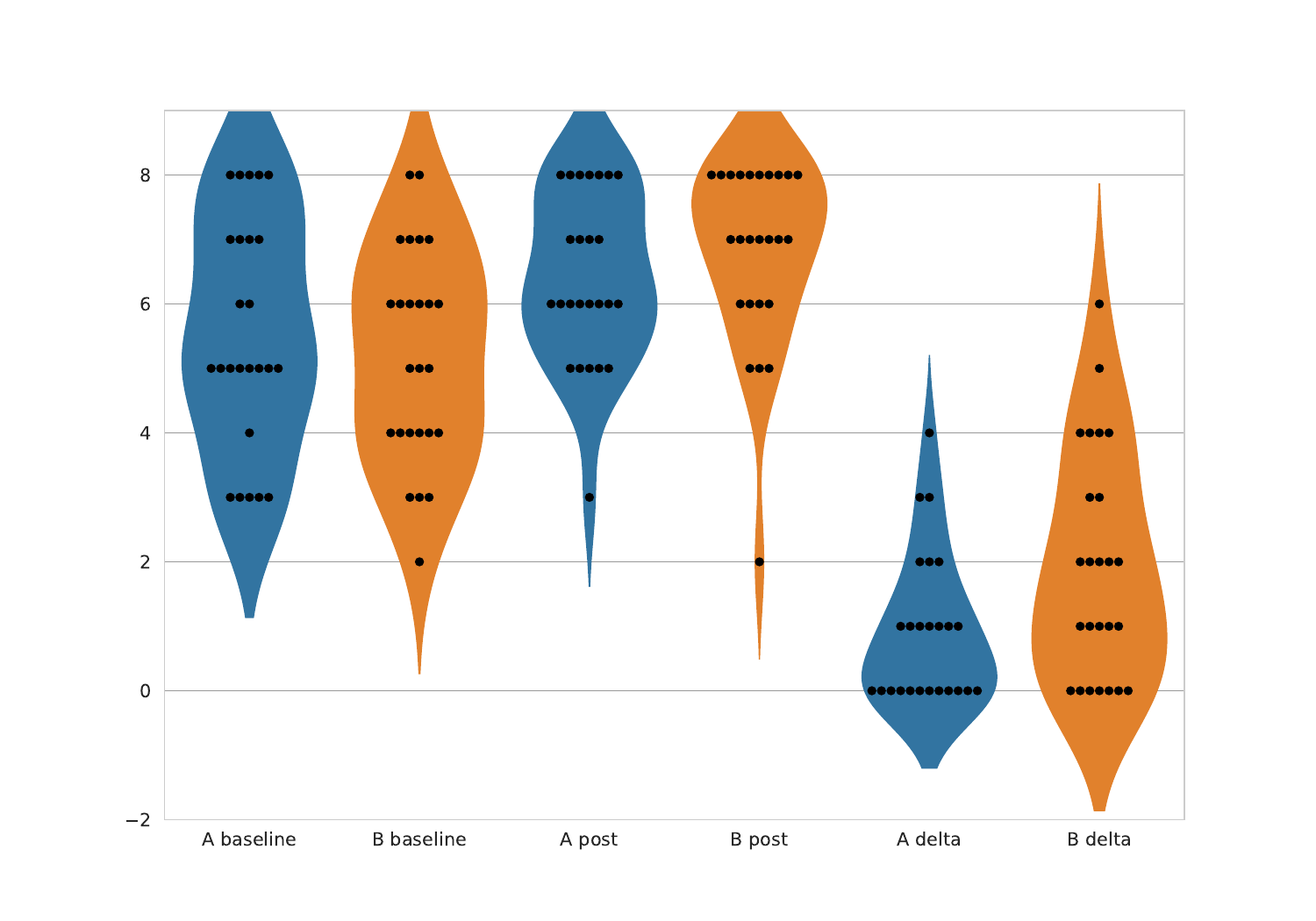}
    \caption{\rev{Violin plots of the distributions for the cancer knowledge scores between groups A and B. The baseline scores for each group follow a similar distribution. Group B scored higher on average than Group A in the post-survey knowledge test (6.4 vs. 6.8); this is further reinforced by the larger average knowledge delta in Group B (0.92 vs. 1.92).}}
    \label{fig:cancer_knowledge}
\end{figure}

\begin{table}[]
\begin{tabular}{lcc}
\toprule
\textbf{Variables} & \textbf{t}     & \textbf{p} \\\midrule
A baseline vs. post-survey knowledge          & 3.46  & 0.002      \\
B baseline vs. post-survey knowledge          & 3.84  & 0.0007     \\
A delta vs. B delta                           & 1.57  & 0.06       \\
Music knowledge vs. delta                     & -0.70 & 0.49       \\
A excitement vs. B excitement                 & 0.99  & 0.32       \\
A valence vs. B valence                       & 4.20  & 0.0001     \\
Cancer experience affect valence?            & 0.11  & 0.92       \\
Cancer experience affect baseline knowledge? & 1.12  & 0.28       \\
Cancer experience affect delta?              & -1.14 & 0.27 \\\bottomrule      
\end{tabular}
\caption{\rev{T-test scores between values. The letters denote the groups that the values belong to. The t-test between music knowledge and knowledge gain was conducted for only group B, as only their music knowledge was assessed. The t-tests between cancer experience and various variables were calculated for all participants; participants were split into groups based on if they answered yes or no to the question.}}
\label{table:t_test}
\end{table}

\rev{We began our analysis by seeking answers to \textbf{Q1}. By using a within-subjects t-test, we determined that both groups A and B scored higher on the cancer-knowledge assessment after finishing the survey (\autoref{table:t_test}, rows 1 and 2). These results indicate a significant gain in knowledge in both groups.}
\rev{Next, we set out to determine whether or not our approach was more effective in educating participants than the article. We first calculated the delta in knowledge survey scores for each group (\autoref{fig:cancer_knowledge}). We then applied a one-tailed between-subjects t-test to yield the result in \autoref{table:t_test}, row 3. While our results are not highly significant (at the p < 0.5 level), the results do imply that our approach is marginally likely (p <.10) to improve cancer literacy levels when compared with a traditional text-based approach.}

\begin{figure}
    \centering
    \includegraphics[width=\columnwidth]{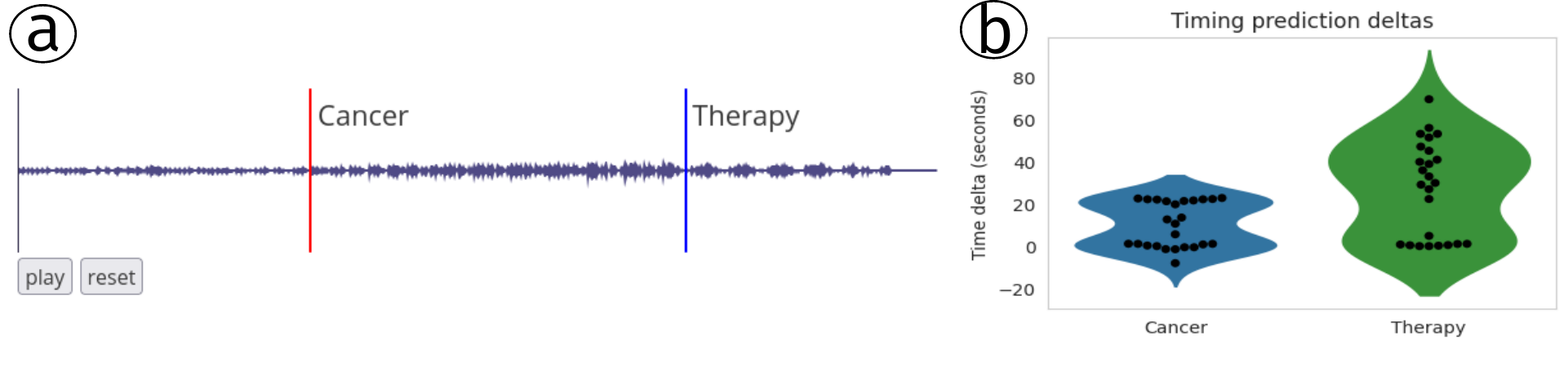}
    \caption{\rev{An illustration of the timing question within the second study. a) is a screenshot of the question's interface, where participants can drag the colored markers to indicate the timestamps where the cancer and therapy started. There are buttons at the bottom to control the playback of the audio. b) describes the distribution of answers to this question. The x-axis is the difference between the actual start of the cancer / therapy and where participants marked the corresponding values.}}
    \label{fig:prediction_delta}
\end{figure}

\rev{To assess the clarity of our approach, we included a task that tested a participant's ability to determine when the cancer started in the piece as well as when the therapy was applied.~\autoref{fig:prediction_delta}a is a screenshot of the interface for the task. Participants would drag the markers corresponding to the timestamps where they believed the cancer and therapy started.~\autoref{fig:prediction_delta}b shows the distance from a participant's response to the actual timestamp when the cancer/therapy began. Participants were able to generally mark when the cancer began (mean = 10.16s, SD = 10.77s), but had trouble marking where the therapy began (mean = 27.35, SD = 22.53s). This is likely due to the fact that the algorithm deletes cancer parts without returning to the original piece. }

\rev{To answer \textbf{Q2}, we examined the relationship between participants' scores in the music knowledge assessment and their cancer knowledge delta. Calculating the Pearson correlation coefficient yielded a value of 0.0765 and a p-value of 0.716, implying a weak positive correlation between music knowledge that does not hold up to significant confidence intervals. This result is also reinforced by the results of the within-subject t-test between the two sets of values (\autoref{table:t_test}, row 4). In other words, this implies that music knowledge has no effect on the efficacy of our approach.}

\rev{To better understand \textbf{Q3}, we examined the differences in emotional impact and excitement experienced. Both approaches received similar ratings in terms of excitement according to the results of the t-test between the two groups (\autoref{table:t_test}, row 5). However, participants generally experienced stronger negative emotions from listening to the music than those participants who read the article alone (\autoref{table:t_test}, row 6).}

\rev{We also examined the relationship between participants' personal experiences with cancer and the emotional impact of our approach. We encoded each participant's response to the question "Do you have personal experience with cancer?" as a binary choice. By comparing the emotional responses (valence) of the group that answered yes to those that answered no, we found that there was no significant relationship between an individual's experience with cancer and the emotional impact that our approach had (\autoref{table:t_test}, row 7).}

\rev{We further examined the relationship between an individual's cancer experiences and their pre-survey cancer knowledge, as well as its bearing on the knowledge delta. The results of the t-tests between the corresponding groups revealed that there was no difference; the cancer experience of an individual does not have any effect on how much they learned from the approach or their prior cancer literacy levels (\autoref{table:t_test}, rows 8 and 9.}).


\rev{\textbf{Short-answer responses}}
\rev{We additionally asked participants in group B to answer some short response questions. These questions can be found in \autoref{appendix:e}.}

\rev{To gauge how well participants understood the approach, we included a question that directly asked participants to explain the analogy back to us. Most responses were terse but accurate descriptions of the approach; the following is a sample:}
\begin{displayquote}
   \rev{``First, you have an original song. This song represents cells, once the song is infected by cancer the original notes from the song start changing. Sometimes the changes are minor and you can still recognize the song, but sometimes it is not. Then, if you apply chemotherapy to the song some "cancer notes" start to turn into rests, these means the previous cell in that position was killed. Some notes do not change, meaning they have some mutations that are resisting therapy.''}  
\end{displayquote}

\rev{Participants in group B largely echoed the sentiments we received from our initial group of participants. 23 participants described the approach as ``interesting'', ``unusual'' or ``unique''. 14 participants referenced learning something or that the approach was informative in some way. 2 participants said that the approach was initially ``strange'' or ``confusing'', but were able to correctly explain the analogy to us.}

\rev{When asked if they preferred reading the article or listening to the music, 23 participants said they preferred the music. Out of the participants who enjoyed the music, 2 said an article was necessary to understand our approach.}
\begin{displayquote}
    \rev{``I think it would be a useful tool explaining cancer to people, it is easier and more pleasant to understand rather that biological books and all big words [sic].''} 
\end{displayquote}

\rev{When asked if they would recommend this approach to others, 21 participants said they would. 3 participants said this approach would be best for children or people with difficulties reading and understanding scientific material. For those who did not say they would recommend the tool, they cited accessibility issues and a need for background knowledge in both music and biology. Moreover, one participant felt that this tool would not help cancer patients:}
\begin{displayquote}
    \rev{``I think I would recommend it for children when learning about DNA mutations. I do not think that cancer patients would benefit from it unless they have good [a] musical ear. It is a great tool; however if I was a cancer patient, would I be focused on how my DNA mutates? I would be more interested in how we could cure it -- as much as this tool is great, it won't cure it. Although again, I think it is amazing for academical [sic] use.''} 
\end{displayquote}

\section{Discussion}
The participants we surveyed found that explaining the mechanisms of cancer through music was effective in reinforcing the concepts explained by the article. The music was able to connect to listeners on an emotional and intuitive level. This can be particularly valuable for listeners who learn in alternative ways, as well as people who may have difficulties reading print media due to stress and other factors. The results also demonstrate that a musical background is not necessary for listeners to be able to appreciate and understand the concepts being illustrated by the musification.

\rev{In terms of broader human-computer interaction research, the results demonstrate that musification is an effective avenue for persuading and educating audiences. Participants spent a fraction of the time with our approach compared with other interactive technologies (such as video games) while demonstrating a similar level of knowledge gain.  Additionally, our approach is more accessible to those with visual and motor impairments which is especially pertinent to those suffering with cancer.}

In the rest of this section, we explore the various suggestions given to us by our participants, the limitations of our study, and directions for future work.

\textbf{Accessibility.} As mentioned in \autoref{sec:results}, participants felt that our approach would be best suited for audiences that do not have strong reading abilities. The participant with cancer felt that patients suffering from the negative cognitive effects of chemo-therapy could also benefit from our approach.~\rev{However, as suggested by one participant, this tool may not be accessible to those with hearing impairments.} No musical background is needed to appreciate the music we generated; this is strongly supported by the work of Bigand~\cite{Bigand.2003.MME, Bigand.2005.MSE}.  Their work demonstrated that people without musical training tend to interpret music similarly to experts. These remarks strongly support our initial goal of educating diverse audiences about cancer. 

\textbf{Musification as an educational tool.} Our approach demonstrates that music has the potential to be a vehicle for explaining scientific concepts. This is supported by the results from the interview, where participants felt that the music conveyed the concept on an intuitive level. The fact that most participants found the audio useful suggests that turning a concept into music lets the resulting piece stand on its own without the need for a complex visualization.

\textbf{Encouragement of early screening.} Our approach inspired some participants to screen early for the disease. This echoes the results of other cancer education approaches~\cite{Acuna.2020.HDV,Jibaja.2000.TIS} where early screening rates were increased as a result of participants interacting with their systems. This suggests that our approach could find success as a public service announcement to inspire audiences to screen for cancer early. A commercial paired with a mutated version of a popular song could possibly be effective.

\subsection{Limitations and Future Work}
\textbf{Context.} One limitation of our approach is the need for context. People need to have a basic knowledge of the abstract concept before understanding the musification. An interesting future direction could be the development of a system that provides an educational walk-through of the mechanisms of cancer, as in our case, which would completely remove the need to rely on traditional print media. This approach would appeal to wider audiences who lack the reading skills necessary to understand complex technical information.

\redout{\textbf{Song choice.} The short duration of ``Twinkle Twinkle Little Star'' was a shortcoming widely observed by our participants. A short piece was selected in an attempt to reduce the complexity of the mutations within so as to avoid overwhelming participants. Moreover, a participant's initial disposition towards a piece could affect the results; choosing a familiar and uplifting piece attempted to mitigate any confusion as well as prevent negative emotional consequences.}

\textbf{The need for an alternative visualization.} We found that participants with a musical background \rev{believed} those without \rev{one} would not understand our approach. \rev{The sheet music in the first study provided details to musically trained participants}\redout{~Since they could interpret the sheet music, they discovered details} that those without formal musical training may have missed. Knowing that the details lie within the sheet music led those participants to believe our approach is not accessible. On the contrary, people without a musical background still relied on those annotations.\rev{This} suggests that people with different levels of music knowledge treated sheet music differently and thus \rev{may have gleaned more information}.\rev{Our approach can be strengthened by providing a visualization that is accessible to those without a musical background.}\redout{It is possible to enable all people to have similar levels of comprehension by increasing the readability of the sheet music visualization.}

\rev{\textbf{Western-centric approach.} Our approach is certainly oriented towards English-speaking audiences that are familiar with ``Twinkle Twinkle Little Star''. We also adopted a western model of musical knowledge, which may have not accurately captured the musical ability of our participants. It would be interesting to apply our approach to popular melodies from different cultures to examine if this impacts its emotional dimension and its overall comprehensibility. Comments from participants indicated that the version of ``Twinkle Twinkle Little Star'' we used was unfamiliar to some; this may have affected how they answered the cancer and therapy questions.}

\rev{\textbf{Visualization and its effect on musical comprehension.} We did not allow participants to see the sheet music in our second study to avoid overwhelming them and tested their ability to determine where the cancer and therapy started within the piece. The majority of participants were not able to properly place the the cancer began within the piece and where the therapy was applied, which contrasts with the claims of the initial participants. This suggests our approach is strongest when coupled with a visualization.}

\redout{\textbf{Small sample size.} Another limitation we acknowledge is the lack of diversity in our sample participant group as well as its small size. Twelve of our thirteen participants were college-educated individuals, which may have affected how they perceived our approach. Due to their high levels of education, participants may have absorbed more information from the print media than the average non-college-educated person. Our approach should be evaluated with a larger, more diverse group of participants.}

\redout{\textbf{Sound effects.} As suggested, the use of sound effects to highlight events occurring during the cell's life-cycle could be explored. Similarly, cancerous notes and themes could be made more dissonant to underline their presence.}

\rev{\textbf{Special interest groups.} A number of participants from both studies suggested that this approach could benefit children as well as individuals who may have difficulty understanding technical scientific materials. A robust future study could examine the effectiveness of our approach when presented to individuals within this group. Additionally, a study exploring the impact of our approach on cancer patients is needed.}

\section{Conclusion}
We designed and implemented an algorithm to musify the mechanisms of cancer. \redout{Our lab study} \rev{We conducted two separate lab studies that demonstrate presenting our approach with an article increases a reader's cancer literacy level more than an article does on its own.}~\redout{showed that coupling our approach with a short article providing context significantly enhanced a listener's understanding of the underlying mechanisms of cancer}. The audio is the primary driving force that helps convey the information and is accessible to people even without any formal musical background. The approach also appeals to listeners on an intuitive and emotional level, which is critical in educational contexts, for people with visual impairments, as well as situations where the listener may not be able to reason to their full potential (e.g., patients undergoing chemotherapy). As such, we believe our approach will be best utilized in educational contexts.\redout{as well as providing cancer patients with another means to interface with their disease.}

\section{Supplementary Material}
We include the questionnaire from our study, an explanatory video, the pre-recorded performances from the study and an additional demonstration of our approach on Pachelbel's Canon in D as supplemental material.
The source code for our program is available at \url{https://github.com/VADERASU/cancer_music}. 

\begin{acks}
We would like to thank Shawn Rupp who implemented an earlier version of this software. We would also like to thank Pauline Davies and Athena Aktipis who helped lead and support the art-science collaborations in the Arizona Cancer Evolution Center, and for the many productive conversations with them over the years about how to understand and represent cancer. 
\end{acks}

\bibliographystyle{ACM-Reference-Format}
\bibliography{main}

\appendix
\section{Likert Scales}
\label{appendix:a}
Before listening:
\begin{itemize}
    \item How much do you think you know about cancer at a cellular level? (From 1 to 5, no knowledge to being expert)
    \item How much do you feel you know about sheet music? (From 1 to 5, no knowledge to being expert)
\end{itemize}
\noindent
After listening:
\begin{itemize}
    \item How much do you think you know about cancer at a cellular level? (From 1 to 5, no knowledge to being expert)
    \item How much do you think the music enhances your comprehension of cancer? (From 1 to 5, no benefit to the most benefit)
\end{itemize}

\section{Interview Questions}
\label{appendix:b}
\begin{itemize}
    \item What was your experience listening to the music?
    \item Was it easy to follow the development and treatment of the cancer? What, if anything, stood out to you during this process?
    \item What can you learn from listening to the music?
    \item How do you feel our approach compares to traditional print media?
    \item Which part is more informative, the sheet music or the audio?
    \item What part do you like/dislike?
    \item If you could change anything about the process, what would you change?
    \item Would you recommend this tool to others? Why?
\end{itemize}

\section{Cancer Literacy Assessment}
\label{appendix:c}
 Questions that ask for the correct mutation had the following multiple-choice answers:
\begin{itemize}
        \item Insertion
        \item Inversion
        \item Deletion
        \item Single nucleotide variation (transposition)
        \item Translocation
\end{itemize}
Items in \textbf{bold} are correct answers.
\begin{itemize}
    \item \textbf{True} or false: Cancer is a genetic disease.
    \item True or \textbf{false}: Chemotherapy always kills cancer cells.
    \item Please choose the statement that is accurate. Cancer cells:
    \begin{itemize}
        \item \textbf{Ignore signals that normally tell cells to stop dividing.}
        \item Infect other cells by attaching to them and inserting their genetic code.
        \item Cannot be removed by the body.
        \item Can only contain one mutation to their genetic code.
    \end{itemize}
    \item Which mutation does this correspond to? Original DNA Sequence: ABCD Mutated DNA: ABABCD \textbf{Insertion}
    \item Which mutation does this correspond to? Original DNA Sequence: ABCD Mutated DNA: ABCF \textbf{Single nucleotide variation (transposition)}
    \item Which mutation does this correspond to? Original DNA Sequence: ABCD Mutated DNA: ACD \textbf{Deletion}
    \item Which mutation does this correspond to? Original DNA Sequence: ABCD Mutated DNA: ACDACBD \textbf{Translocation}
    \item Which mutation does this correspond to? Original DNA Sequence: ABCD Mutated DNA: BACD \textbf{Inversion}
\end{itemize}

\section{Music Knowledge Assessment}
\label{appendix:d}
Items in \textbf{bold} are correct answers.
\begin{itemize}
    \item What is the duration of this note? \textbf{Half-note}
    \item[] \includegraphics[width=4cm,height=2cm]{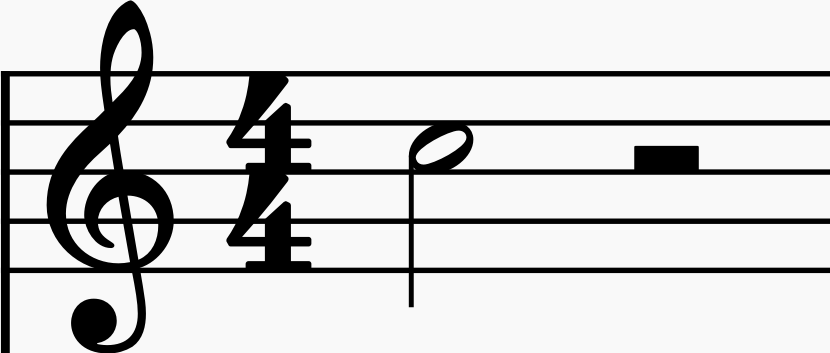}
    \item Please identify the name of this note. \textbf{C}
    \item[] \includegraphics[width=4cm,height=2cm]{figures/half_note.png}
    \item Please identify the name of this note. \textbf{F}
    \item[] \includegraphics[width=4cm,height=2cm]{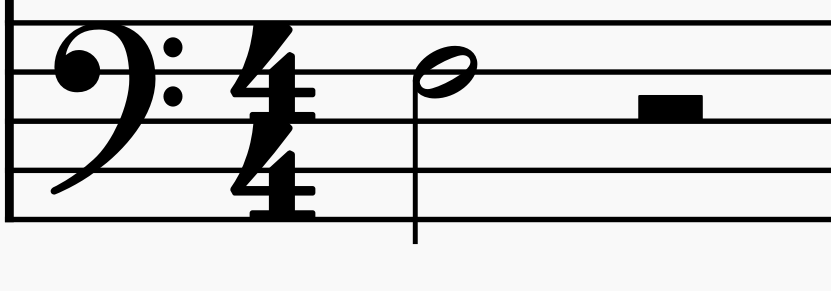}
    \item Please identify this interval. \textbf{Minor sixth}
    \item[] \includegraphics[width=4cm,height=2cm]{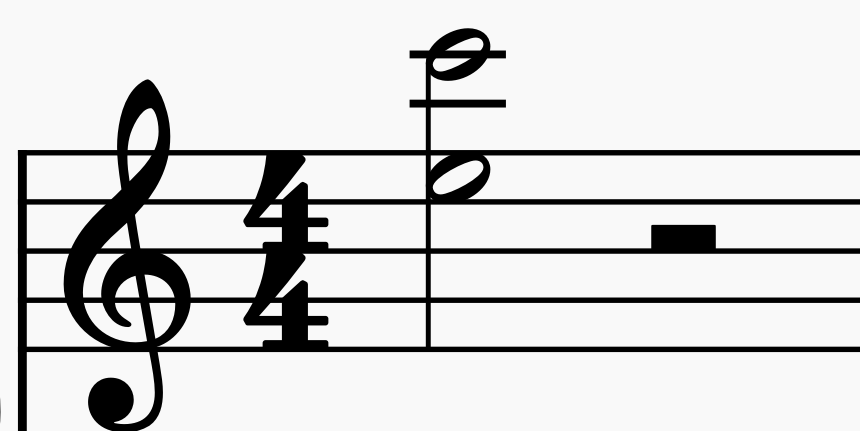}
    \item Please identify this chord. \textbf{F major}
    \item[] \includegraphics[width=4cm,height=2cm]{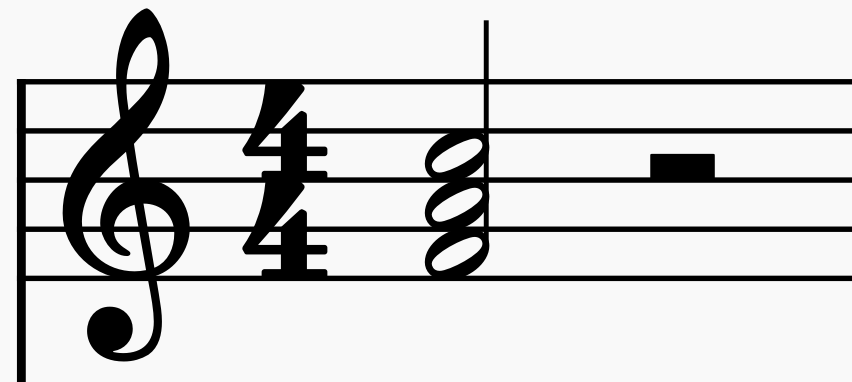}
\end{itemize}

\section{Second study interview questions}
\label{appendix:e}
\begin{itemize}
    \item In your own words, please explain the analogy we used to generate the music back to us.
    \item What was your experience as a whole?
    \item What did you learn from listening to the mutated music?
    \item How does this approach to teaching people about cancer compare to the article you read?
    \item Would you recommend this tool to others? Why?
\end{itemize}
\end{document}